\documentclass{article}
\usepackage{icmc2020template}
\usepackage{times}
\usepackage{ifpdf}
\usepackage{soul}
\usepackage[english]{babel}
\def\papertitle{Network Modulation Synthesis: New Algorithms for Generating Musical Audio Using Autoencoder Networks}
\def\firstauthor{Jeremy Hyrkas}
\def\secondauthor{Second Author}
\def\thirdauthor{Third Author}

\newif\ifpdf
\ifx\pdfoutput\relax
\else
   \ifcase\pdfoutput
      \pdffalse
   \else
      \pdftrue
  \fi
\fi

\ifpdf 
  \usepackage[pdftex,
    pdftitle={\papertitle},
    pdfauthor={\firstauthor, \secondauthor, \thirdauthor},
    bookmarksnumbered, 
    pdfstartview=XYZ 
   ]{hyperref}

  \usepackage[pdftex]{graphicx}
  \graphicspath{{./figs/}}
  \DeclareGraphicsExtensions{.pdf,.jpeg,.png}

  \usepackage[figure,table]{hypcap}

\else 
  \usepackage[dvips,
    bookmarksnumbered, 
    pdfstartview=XYZ 
  ]{hyperref}  

  \usepackage[dvips]{epsfig,graphicx}
  \graphicspath{{./figures/}}
  \DeclareGraphicsExtensions{.eps}

  \usepackage[figure,table]{hypcap}
\fi

\hypersetup{
    colorlinks,%
    citecolor=black,%
    filecolor=black,%
    linkcolor=black,%
    urlcolor=black
}

\title{\papertitle}

\oneauthor
   {\firstauthor} {University of California San Diego  /  New York University \\ %
     {\tt \href{mailto:jhyrkas@ucsd.edu}{jhyrkas@ucsd.edu}}}

\begin{document}
\capstartfalse
\maketitle
\capstarttrue
\begin{abstract}
A new framework is presented for generating musical audio using autoencoder neural networks. With the presented framework, called network modulation synthesis, users can create synthesis architectures and use novel generative algorithms to more easily move through the complex latent parameter space of an autoencoder model to create audio. 

Implementations of the new algorithms are provided for the open-source CANNe synthesizer network, and can be applied to other autoencoder networks for audio synthesis. Spectrograms and time-series encoding analysis demonstrate that the new algorithms provide simple mechanisms for users to generate time-varying parameter combinations, and therefore auditory possibilities, that are difficult to create by generating audio from handcrafted encodings. \end{abstract}

\section{Introduction}\label{sec:introduction}

Neural networks for audio generation have recently been a hotbed for research in computer music, music information retrieval, and artificial intelligence research communities. Early research into audio synthesis using neural networks focused on generating speech audio, with pioneering models WaveNet \cite{wavenet} and SampleRNN \cite{samplernn} making strong breakthroughs using deep neural networks for speech synthesis. 

Following efforts in speech synthesis, music researchers have targeted deep learning models as a technology for musical audio generation and manipulation. WaveNet, combined with bidirectional recurrent neural networks, has been used to synthesize musical sounds based on user-selected points in a timbral space \cite{timbre}. Neural networks have been used to create bizarre mashups of existing music \cite{blackmetal} and hallucinate new music based on self-modifying training data \cite{dreaming}. Following image processing examples, researchers have attempted auditory style transfer \cite{style}. While these and other models can be used to create interesting music, the complexity of the models result in lengthy training and generation times.

Recently, autoencoder networks have been used for audio generation and modification to address this issue. Autoencoders tend to have smaller architectures than complex convolutional and recurrent neural networks. Depending on the specific architecture, some autoencoders can generate audio in or near real-time, and have the additional benefit of allowing users to modify the generated audio more easily than other networks. WaveNet again provided an early example, as it was augmented with an autoencoder that allowed users to interpolate between instrument encodings to combine timbral elements of each instrument \cite{wavenet2}. However, the decoder was not fast enough for compositional purposes, and the authors chose to release the model as software with pre-defined latent representations that are interpolated, disallowing the user to explicitly manipulate the latent parameter space. The model leveraged in this work, the CANNe synthesizer \cite{canne}, allows users to create audio by specifying only eight latent parameter values and works in near real-time.

Naturally, academic musicians and music researchers were the first to explore neural network musical synthesis. However, deep learning technology is ubiquitous across all aspects of society, and experimental musicians from beyond the academic space have begun incorporating neural networks into their art. Holly Herndon’s 2019 album \emph{PROTO} \cite{herndon} uses a neural network agent trained on Herndon’s voice as a co-vocalist among a full musical mix, while experimental duo Emptyset created an album made entirely from sounds generated by a model trained on their music on 2019’s \emph{Blossom} \cite{emptyset}. With growing interest from electronic artists in composing with musical neural networks, there is ample opportunity for researchers to explore methods for musical generation that are accessible to musicians in both parameter control and time to music creation. Because autoencoders can excel at both of these metrics, this work explores new algorithms for music generation using autoencoders that allow users to perform semantically simple parameter changes that move through the network’s complex latent parameter space.

\section{Background and motivation}\label{sec:background}

\subsection{CANNe synthesizer model}\label{sec:canne}

\begin{figure*}[ht]
\includegraphics[scale=0.3]{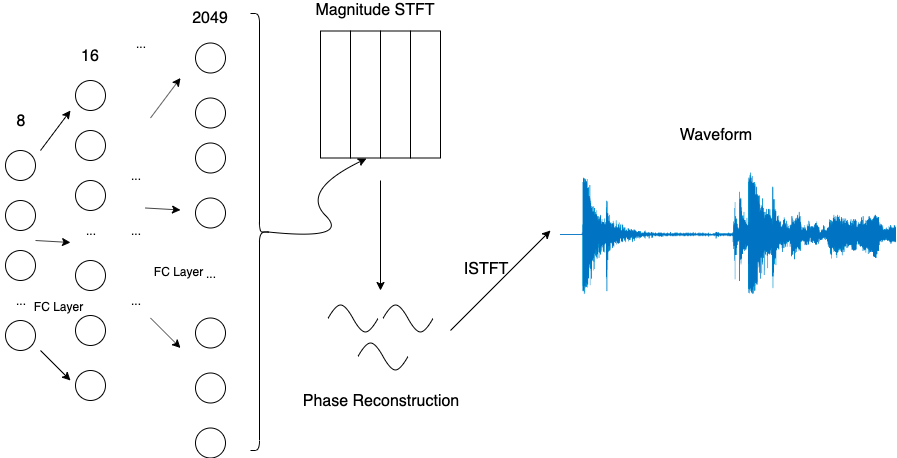}
\centering
\caption{The decoding section of the CANNe autoencoder. The user provides the values of eight latent parameters, which are passed through the decoder (comprising of increasingly large fully connected layers) until a final output layer of size 2049 has its output reflected to become the magnitude of a 4096-point Discrete Fourier Transform (DFT). The magnitude DFT is inserted into a Short-time Fourier Transform (STFT) buffer. The STFT is augmented with a phase-reconstruction algorithm, and an inverse-STFT results in time-domain audio.}
\label{fig:decoder}
\end{figure*}

Colonel et al.’s method for generating musical audio, the CANNe (Cooper’s Autoencoding Neural Network) synthesizer \cite{canne}, is directly leveraged in the network modulation synthesis software presented here; as such, it is instructive to review the method as presented before discussing modifications and extensions.

An autoencoder network is a neural network used to learn an encoding and decoding method for a data set. The goal is often to learn an efficient encoding for input data, or to learn some fundamental parameters of a type of data with a dimensionality much smaller than the original. Autoencoder networks are split into two sections: the encoder section and the decoder section. In a standard deep neural network autoencoder, the encoder section takes as input the training data and learns using one or more fully connected layers with decreasing size. The decoder section takes as input the output values of the final encoding layer and learns one or more fully connected layers of increasing size. The final layer of decoding is of the same dimensionality as the training set, and the overall network is optimized to learn an identity function.

In the CANNe synthesizer architecture, the network takes as input the magnitude of the first 2049 bins from a 4096-point Discrete Fourier Transform (DFT) and learns 15 fully connected layers. The layers of the encoder section decrease roughly by powers of two (with the exception of the first layer of size 1000 instead of 1024) until the middle layer, made up of eight neurons. The layers of the decoder section double in size, finally resulting in an output layer of size 2049. The training corpus contains 91,000 STFT frames from 5-octave C-major scales played on a MicroKORG synthesizer across 80 patches. Once trained, a user can manufacture outputs from the innermost layer (i.e. create an encoding), causing the decoder to create a magnitude frame. A state-of-the-art phase reconstruction algorithm \cite{phase} is used to convert the magnitude frames to complex-valued STFT frames that can be inverted to a time-domain audio signal. The small network size coupled with the efficient phase reconstruction algorithm allows the network to generate audio in near real-time. The encoding-to-audio generation process is visualized in Figure \ref{fig:decoder}.

\subsection{Encoder artifacts and parameter choice}\label{sec:parameters}

Interviews with artists such as Herndon reveal that part of the appeal of composing with statistical models are the limitations and resulting artifacts in the generated audio \cite{herndon_interview}. One benefit of the CANNe architecture’s choice to allow users to directly choose an encoding is that users can find encodings that were rare or not present in the original data set, resulting in the model producing glitchy and artifact-filled audio. Additionally, autoencoder models with such a small encoding are unlikely to achieve 100\% training and test accuracy, meaning that deficiencies in the decoder are made explicit in the generated audio, which can fall into an uncanny valley relationship with the training set. Exploring these uncanny audio examples is of interest to artists such as Emptyset \cite{emptyset_interview}.

But what of the encoder? The encoding-to-audio generation scheme only allows users to generate audio using the decoder portion of the network after training, so artifacts and deficiencies in the encoder cannot be exploited as musical aesthetic. Additionally, the CANNe architecture features an encoder and decoder with the same number of layers with similar but not identical sizes, meaning that the decoding function is likely not an exact mathematical inverse of the encoding. Other autoencoder models may follow this architecture choice, as it can lead to higher accuracy on some problems \cite{asym_autoencoder}. It would be beneficial to expose the encoder model for musical synthesis to expand the sonic possibilities of the network.

One option is to allow the user to create inputs to the full autoencoder model. However, this solution is unacceptable for two reasons. First and most obviously, the size of the autoencoder’s input layer (2049 data points per audio frame) makes hand-crafting inputs intractable. Second and more subtly, the model should see input data that is similar to its training corpus to produce output beyond chaotic noise. Therefore, the methods presented here aim to allow users to utilize the full autoencoder model to generate audio while limiting the number of tunable parameters.

Lastly, Section \ref{sec:latent_params} will demonstrate that similar sounding audio can be generated from dissimilar encodings. Due to the immense amount of complex timbral information stored in such a small encoding, small changes to the latent parameter space can drastically alter the audio, and drastic changes can theoretically result in near-identical audio. Therefore, an added benefit of the methods presented is the ability for users to easily make subtle changes to generated audio by changing one parameter, as opposed to extensively searching the latent parameter space.

\section{Methods}\label{sec:methods}

\subsection{Network Modulation Framework}\label{sec:network_mod}

\begin{figure*}[ht]
\includegraphics[scale=0.4]{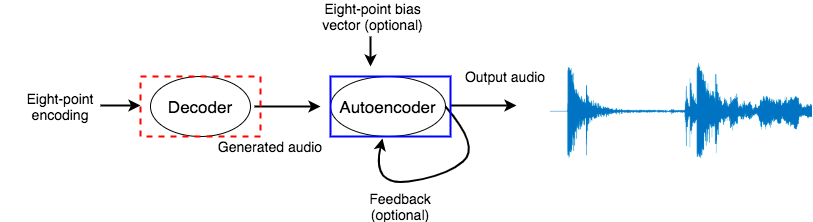}
\centering
\caption{A basic architecture for network modulation synthesis is depicted above. Audio created from the decoder (or modulator network, marked by the red dotted box) is used as input to the full autoencoding model (or carrier network, marked in the blue solid-lined box). In addition to the eight latent parameters necessary for the decoder, a user may optionally specify a bias vector for the autoencoder and a feedback percentage.}
\label{fig:networkmod}
\end{figure*}

Though the CANNe synthesizer operates by creating 8-point encoding values for the decoder, the network is fundamentally trained to generate outputs based on magnitude STFT frames. A natural expansion to the synthesis framework would be to probe the network using STFT data created from time-domain audio, thus taking advantage of the timbral qualities of the encoder, which was previously unused in audio generation. To do this, we have expanded the CANNe codebase with a utility function to generate one magnitude frame by either 8-point values to probe the decoder or 2049-point values to probe the entire autoencoder. If the network had perfectly learned an encoding scheme for the training corpus, running a full encoding/decoding prediction would have no effect on the generated audio if it closely resembles a training example. Luckily, the model is imperfect, and the audio contains new timbral artifacts after it is passed through the network.

The trained model made available by Colonel et al. only learns bias vectors for the input and innermost layer, as the authors found that additional bias vectors raised the noise floor of the generated audio. However, it is possible to reassign the bias variables of the neural network. In keeping with the spirit of the original CANNe framework, audio generated using the full autoencoder can be additionally controlled by an 8-point vector. In the original CANNe framework, the 8-point vector is used as the output of the innermost layer; in this method, the 8-point vector is used as the bias vector for the layer prior to the innermost. By allowing a user to tweak the bias vector in the final encoding layer, the generated audio can still be adjusted even if the input audio is kept constant.

While an autoencoder network can theoretically be probed with an STFT frame from any audio, statistical models work best on data that closely resemble their training corpus. One method for ensuring this similarity is to use audio generated in the original CANNe framework as input audio. In this framework, a user can create audio by constructing an 8-point input vector to probe the decoder and use the STFT frames of the generated audio (optionally alongside an 8-point bias vector) to predict new audio frames. This framework has a metaphorical similarity to frequency modulation synthesis, or FM synthesis \cite{fm}, in which the instantaneous frequency of an oscillator (the carrier) is controlled by the output of another oscillator (the modulator). Following this metaphor, the original CANNe network using the decoder for synthesis can be thought of as a modulator network, with its output being used as input to the full autoencoder, or carrier network. Later, it will be shown that more complex architectures can be used where a network may act as both a modulator and carrier network, but for now, we will define a modulator network as a network generating audio via the decoder model, and a carrier network as a network generating audio via a full network prediction. This general algorithm is called \emph{network modulation synthesis}. A basic modulator/carrier pair is demonstrated in Figure \ref{fig:networkmod}.

To further borrow from FM synthesis design, feedback can be used in network modulation synthesis. A carrier network can add a portion of the magnitude frame from its previous prediction to the incoming magnitude frame from the modulator network to form a new input. As in other audio algorithms, the user can specify the feedback amount through a feedback variable valued between 0.0 and 1.0. Section \ref{sec:spectral} shows that feedback offers users a simple, intuitive parameter that can subtly or drastically alter the generated audio.

\subsection{Implementation}\label{sec:architecture}

\begin{figure*}[ht]
\includegraphics[scale=0.4]{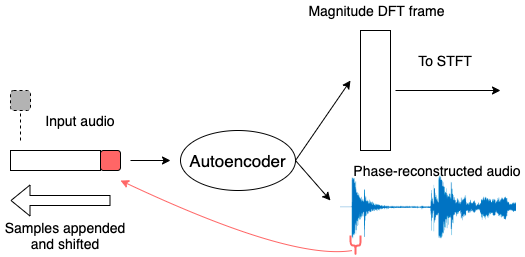}
\centering
\caption{Predictive feedback algorithm for autoencoders. Input audio is used to generate an STFT frame, which is also converted to time-domain audio after phase reconstruction. The first five samples of the input audio are dropped and the first five samples of generated audio are appended to the end of the input audio. This process continues iteratively until the desired amount of STFT frames are generated.}
\label{fig:predictive_feedback}
\end{figure*}

For ease of use, a compositional utility class was created in the CANNe codebase. The class manages the underlying steps to generate audio for the modulator and carrier algorithms. In the case of the modulator algorithm, a user specifies the 8 latent parameters to probe the decoder and the desired length $L$ in seconds of the specified audio. The compositional utility class generates the amount of magnitude frames necessary for $L$ seconds of audio, creates an STFT using phase reconstruction \cite{phase}, and performs an inverse-STFT to output the musical audio at 44.1 kHz. For the carrier algorithm, the user provides input audio as mono audio data sampled at 44.1 kHz, and optionally a bias vector and feedback amount. The function performs an STFT on the input audio, takes the magnitude of the STFT and normalizes frames so that the maximum frequency bin value is 1, matching the normalization of the CANNe training data. Every STFT frame from the input becomes a predictive input for the autoencoder, and every predicted output fills a corresponding STFT magnitude frame. The output is converted to time domain audio identically to the modulator algorithm. Performing STFT and ISTFT operations between every network was chosen for ease of implementation and could result in unnecessary computation; however, it does allow predicted audio to be further modified based on artifacts from the phase reconstruction algorithm.

To add flexibility to the compositional utility class, users may specify the 8-point decoder or bias variables as either a length 8 vector, or as an $N\times8$ matrix. In the case of the latter, each row of the matrix is used to generate one magnitude frame of the output. This allows a user to change the values of the latent parameters over time, which often creates the most interesting timbral effects.

While the compositional utility class provides somewhat higher level access to the underlying neural network, a more sophisticated abstraction is necessary to remove the manual process of generating audio in various synthesis architectures. This abstraction comes through the Architecture class.

A synthesis Architecture contains one or more instances of an OscillatorNetwork class. The first instance is the root, a modulator network. Additional instances can be added; each further network is a carrier network, with the user specifying which existing network acts as its modulator. In the current implementation, there can be only one root, but each carrier network can act as a modulator for another carrier; thus, every architecture takes the form of a tree data structure. Each OscillatorNetwork instance has its own parameters that can be set between audio generation. To generate audio, the user specifies the length of audio desired and the architecture will create audio at each network, running through the tree and returning a list of audio from the leaves of the network. Underneath the Architecture and OscillatorNetwork abstractions, only one instance of the neural network model is instantiated.

\subsection{Signal processing considerations}

In addition to the latent parameters for each network, the synthesis framework allows users to specify a pitch shift and amplitude envelope to apply to generated audio. Pitch shifting is performed using an implementation available through the Python package LibROSA \cite{librosa}. Envelopes can be specified as either an exponential decay, or as simplified attack, decay, sustain and release (ADSR) parameters. In the simplified schema, the user should provide an attack time, attack level, decay time, sustain level, and release time. There is no sensible notion of “releasing” when generating audio from a neural network, so the release time is applied backwards from the end of the signal. 

While pitch shifting and enveloping are far from novel in audio synthesis, they must be applied differently in generative models than in traditional signal processing. In predictive models, the input data should be kept as close as possible to the format of the training data, which almost always entails input normalization; in other words, the normalization necessary for prediction nullifies the envelope of the input audio. Similarly, pitch shifting input audio may result in audio with pitch classes that were not present in the training data. As a result, enveloping and pitch shifting are applied only to the final audio generated at each leaf of the architecture. This limitation is markedly different from FM synthesis, where enveloping and pitch altering can be performed at each oscillator in the architecture.

\subsection{Predictive Feedback}\label{sec:predictive_feedback}

Framing network modulation synthesis as analogous to frequency modulation synthesis led directly to the implementation of a feedback mechanism similar to those found in signal processing. However, predictive models are fundamentally different systems to signal processing systems, and the differences in function allows for distinct methods for audio generation. In the case of feedback, audio can be iteratively created by altering the input audio slightly using predicted output audio. 

Pfalz et al. suggest an algorithm for audio generating using long short-term memory units (LSTMs) called \emph{dreaming} \cite{dreaming}. The dreaming algorithm is suited for neural networks trained to take a sequence of input audio and predict the subsequent audio. New audio is dreamt by iteratively predicting new audio based on an input seed, dropping the first section of the seed and appending the predicted audio and re-predicting. The auditory effect of dreaming is highly dependent on the training problem, as the model may be trained to predict the next frame of audio in the time- or frequency- domain, or, in the most extreme case, the next sample of time-domain audio.

The underlying CANNe network is a simpler model than the LSTMs used in dreaming, and the training problem is incompatible as the autoencoder attempts to learn the identity function, as opposed to learning the next data point in a sequence. However, ideas from Pfalz et al.’s algorithm have been adapted here in a method referred to as \emph{predictive feedback}. When a carrier network performs predictive feedback, only the first frame of audio from a modulator network is used as input. To generate each frame of output audio, the input and predicted audio are first converted to time-domain audio. The first five samples of the input audio are then dropped, and the first five samples of predicted audio are appended to the input audio. The new input audio is then converted to the frequency domain for prediction. Figure \ref{fig:predictive_feedback} illustrates this process.

\begin{figure*}[ht]
\includegraphics[scale=0.5]{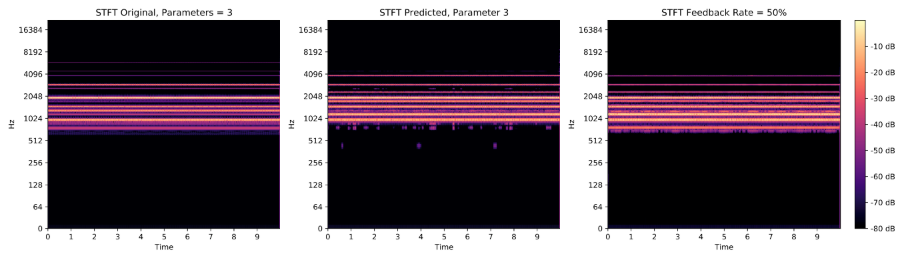}
\centering
\caption{Spectral effect of network modulation synthesis. The left figure demonstrates audio generated from the vanilla CANNe synthesizer with an encoding of all 3.0 values. The middle figure demonstrates the same parameters in a basic network modulation synthesis architecture with no bias or feedback; the right figure demonstrates the same process with 50\% feedback. The spectrograms differ, demonstrating that network modulation synthesis allows the user a simple method to explore sonically-related areas of the latent parameter space.}
\label{fig:netmod_filter}
\end{figure*}

The choice of five samples comes from subjective testing. When altering the input audio by only one sample, the frequency content is changed so imperceptibly that the resulting audio is indistinguishable from audio predicted using only the original input. When the number of samples becomes too large, noticeable discontinuities in the input audio lead to poor audio prediction; in fact, the extreme case of replacing the input audio entirely with the predicted audio is equivalent to using traditional feedback with a feedback variable of 1.0. Five samples seems to allow predictive feedback to smoothly migrate from the original input without noticeable discontinuities, but further exploration of this method and its effects on prediction could lead to a more robust hyperparameter choice. 

To perform predictive feedback, a carrier network can be instantiated in predictive feedback mode. Unlike standard modulator and carrier networks, networks in predictive feedback mode have no other variables. Section ~\ref{sec:spectral} demonstrates that predictive feedback allows users to create dynamically changing audio without the need to specify time-varying parameters.

\section{Analysis}\label{sec:analysis}

\subsection{Spectral analysis of network modulation}\label{sec:spectral}

Figure \ref{fig:netmod_filter} demonstrates a spectral example of network modulation synthesis. As predicted, the encoding and decoding portions of the autoencoder model have different artifacts as a result of imperfect training and test accuracy. As a result, the audio generated from the vanilla CANNe model is different from the same audio refiltered through the autoencoder, even with no user-provided bias parameters. The audio is again changed through the use of feedback. This result confirms that even the most basic network modulation architecture can allow users to explore related latent parameter spaces without the need to specify new encoding parameters. The addition of bias parameters, while not depicted, trivially changes the parameter space as the bias vector is added directly to the final encoding.

\begin{figure}[ht]
\includegraphics[scale=0.4]{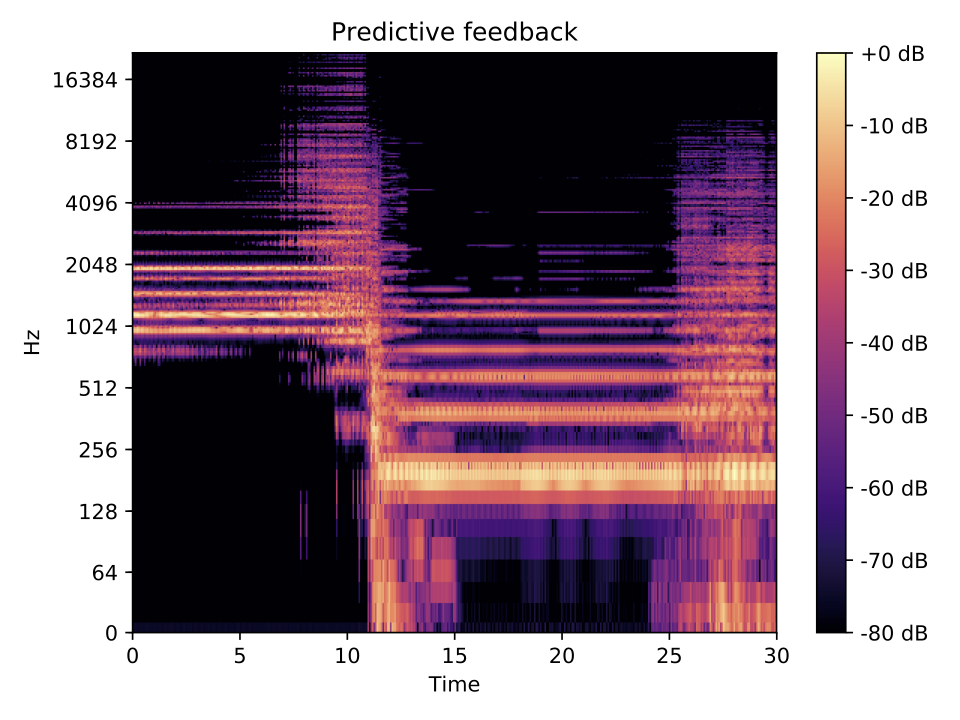}
\centering
\caption{Spectral effect of predictive feedback. As the input audio is modified by the predicted audio, the network moves through different regions of the latent parameter space and the audio slowly evolves. This process is achieved with no parameter modulation.}
\label{fig:predfeed_spect}
\end{figure}

Predictive feedback is depicted in Figure \ref{fig:predfeed_spect}. The process of predictive feedback allows users to produce audio that constantly moves through the parameter space without needing to specify time-varying parameter choices. 

\subsection{Latent parameter analysis}\label{sec:latent_params}

\begin{figure*}[ht]
\includegraphics[scale=0.4]{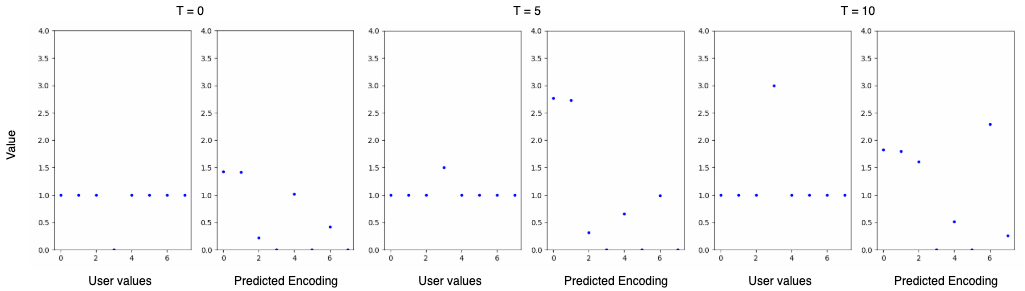}
\centering
\caption{Clips from a parameter sweep in a basic network modulation synthesis architecture. In a modulator network, parameter 3 is swept from 0.0 to 3.0 over ten seconds and every other parameter is set to 1.0. The left side of each time stamp shows the user-provided encoding parameters. The right side of each time stamp shows the predicted encoding from a carrier network. This example illustrates the complex relationship between latent parameters, and how the network modulation algorithm allows a user to move through the latent parameter space in a manner that would be difficult to code by hand.}
\label{fig:paramsweep}
\end{figure*}

A key hypothesis underlying the motivation for network modulation synthesis is that similar sounding audio can result from differing latent parameter choices. This hypothesis is tested and confirmed in Figure \ref{fig:paramsweep}. In a basic modulator and carrier network pair, seven parameters of the modulator network are fixed while one is slowly swept from 0.0 to 3.0, resulting in slowly morphing audio. The generated audio was passed through the encoder and the encoding was recorded. As demonstrated in the figure, all but two of the eight encoding parameters move as only one parameter is changed in the modulator network. In the animated results of the experiment, it is also clear that the encoding values wobble and modulate as the modulator network moves through its parameter space. This result is key to the algorithm's effectiveness, as it demonstrates that even though all sounds created in the experiment can theoretically be achieved by feeding the correct series of encodings to the decoder, the network modulation synthesis algorithm can be used to easily move through parameter spaces that would be near-impossible to generate by hand.

\section{Discussion}\label{sec:discussion}

The network modulation synthesis framework contains several new algorithms for musical audio generation using autoencoder networks. The core synthesis architecture allows for additional latent parameter choices in the form of a bias vector, as well as feedback options in the form of a traditional feedback ratio or the new predictive feedback algorithm. While the basic modulator / carrier pair has been focused on thus far, complex architectures fanning out into several output channels are possible.

The algorithms presented can be used on most autoencoder architectures. Implementations are provided in the CANNe synthesizer codebase because of its open-source availability and quick generation time. Additional work outside the scope of this paper has included several compositions composed using the expanded codebase and a new graphical user interface to allow users to generate audio using time-varying parameters by drawing parameter paths. These compositions and expansions further demonstrate the viability of the CANNe synthesizer and the network modulation framework as compositional tools. 

Whether or not a community builds around the expanded CANNe codebase, autoencoder networks continue to be a thriving research topic for musical audio generation. Future commercial or open-source musical autoencoders can benefit by including the algorithms presented here. As commercial interest in AI-generated music continues to grow, new algorithms and user control methods like those presented here offer novel possibilities for experimental compositions and performance systems.

Audio clips accompanying the analyses presented here can be found at \href{https://jeremyhyrkas.com/ICMC2020}{https://jeremyhyrkas.com/ICMC2020}, and all the presented software is open-source and available at \newline\href{https://github.com/jhyrkas/canne_synth}{https://github.com/jhyrkas/canne\_synth}.

\bibliography{paper}

\end{document}